\documentclass[]{spie}  
 
\usepackage{amsmath,amsfonts,amssymb}
\usepackage{graphicx}
\usepackage[colorlinks=true, allcolors=blue]{hyperref}
\usepackage{float}

\title{The High-Speed FPGA Readout System for the Advanced X-ray Imaging Satellite (AXIS)}

\author[a]{Declan O'Neill}
\author[b]{Jill Juneau}
\author[a]{Peter Orel}
\author[a]{Sven Herrmann}
\author[b]{Gregory Prigozhin}
\author[b]{Eric D. Miller}
\author[a,c,d]{Steven W. Allen}
\author[b]{Marshall W. Bautz}
\author[a]{Tanmoy Chattopadhyay}
\author[e]{Kevan Donlon}
\author[b]{Robert Goeke}
\author[b]{Catherine E. Grant}
\author[b]{Beverly LaMarr}
\author[e]{Christopher Leitz}
\author[a,d]{R. Glenn Morris}
\author[a,c]{Abigail Y. Pan}
\author[a,c]{Tonya L. Peshel}
\author[a]{Artem Poliszczuk}
\author[a,c]{Haley R. Stueber}
\author[e]{Keith Warner}

\affil[a]{Kavli Institute for Particle Astrophysics and Cosmology, Stanford University, 452 Lomita Mall, Stanford, CA 94305, USA}
\affil[b]{MIT Kavli Institute for Astrophysics and Space Research, Massachusets Institute of Technology, 70 Vassar St, Cambridge, MA 02139, USA}
\affil[c]{Department of Physics, Stanford University, 382 Via Pueblo Mall, Stanford CA 94305, USA}
\affil[d]{SLAC National Accelerator Laboratory, 2575 Sand Hill Road, Menlo Park, CA 94025, USA}
\affil[e]{MIT Lincoln Laboratory, 244 Wood St building 1324, Lexington, MA 02421, USA}

\authorinfo{Further author information: Send correspondence to D. O'Neill\\D. O'Neill: E-mail: declano@stanford.edu}

\begin{document} 
\maketitle


\begin{abstract}
The Advanced X-ray Imaging Satellite (AXIS) is a Probe-class mission concept designed to deliver arcsecond spatial resolution, high-sensitivity spectral imaging across the 0.3–10 keV band. The X-ray Astronomy and Observational Cosmology (XOC) Group at Stanford, in collaboration with the MIT Kavli Institute (MKI) and MIT Lincoln Laboratory (MIT-LL), is developing the AXIS X-ray camera, including both the detector and the front-end readout electronics required to meet the mission’s demanding performance goals. The telescope’s focal plane detector consists of  four 1440×1440 pixel charge-coupled devices (CCDs) developed by MIT-LL, each featuring 16 parallel output channels. These outputs are amplified by a high-speed, low-power, low-noise application-specific integrated circuit (ASIC) – the Multi-Channel Readout Chip (MCRC) – developed at Stanford. Following amplification, the analog signals are digitized and processed to construct a pixel array, prior to event detection. Here, we present the field-programmable gate array (FPGA) architecture developed to enable high-speed, parallelized readout of these CCD channels. The FPGA samples 16 analog-to-digital converter (ADC) channels at 50 MHz, performs preprocessing of pixel data, which is then streamed via User Datagram Protocol (UDP) over a 1 Gb Ethernet link to the back-end system for event reconstruction. Our design demonstrates the goal readout performance for AXIS (20 frames per second) and provides a framework for future high-throughput X-ray observatories.
\end{abstract}


\keywords{AXIS, FPGA, CCD, X-ray detector, readout electronics, instrumentation}


\section{INTRODUCTION}

One of the priorities identified in the 2020 Decadal Survey was a medium-scale X-ray mission to complement the European Space Agency's Athena observatory.\cite{astro2020} The Advanced X-ray Imaging Satellite (AXIS) is a Probe-class mission concept developed in response to NASA's 2023 Astrophysics Probe solicitation. AXIS is designed to advance the capabilities of X-ray astronomy through the combination of large effective area, high angular resolution, and high-sensitivity imaging\cite{reynolds23}, dramatically improving our ability to identify and characterize faint, distant sources. The AXIS focal plane camera is being developed by a collaboration between Stanford's X-ray Astronomy and Observational Cosmology (XOC) Group, MIT Kavli Institute (MKI), and MIT Lincoln Laboratory (MIT-LL). To achieve the wide field of view and high spatial resolution required for the mission, MIT-LL has fabricated prototype large-format (1440×1440 pixel) charge-coupled devices (CCDs).\cite{chris25-ccid100} These large CCDs introduce a key challenge: reading out the sensors at frame rates fast enough to avoid photon pile-up.\cite{lumb00_pileup_xmm}

This paper describes the field-programmable gate array (FPGA)-based front-end electronics (FEE) architecture developed to enable high-speed, parallel readout of the AXIS CCDs. In this design, the CCD output channels are amplified by a dedicated Multi-Channel Readout Chip (MCRC) application-specific integrated circuit (ASIC),\cite{herrmann20_mcrc,porelMCRCspie2024} digitized by multi-channel high-speed analog-to-digital converters (ADCs), and streamed as pixel data to the back-end electronics (BEE) for event reconstruction. We focus on the digital architecture and firmware that (i) ingests ADC samples, (ii) performs correlated double sampling (CDS) and pixel formation, (iii) buffers and packetizes pixel data, and (iv) streams the resulting data over 1 Gbit Ethernet using User Datagram Protocol (UDP). We also present results from initial testing using a representative single-ADC test platform.


\section{AXIS CAMERA AND FEE}

\begin{figure}[t!]
    \centering
    \includegraphics[width=.6\linewidth]{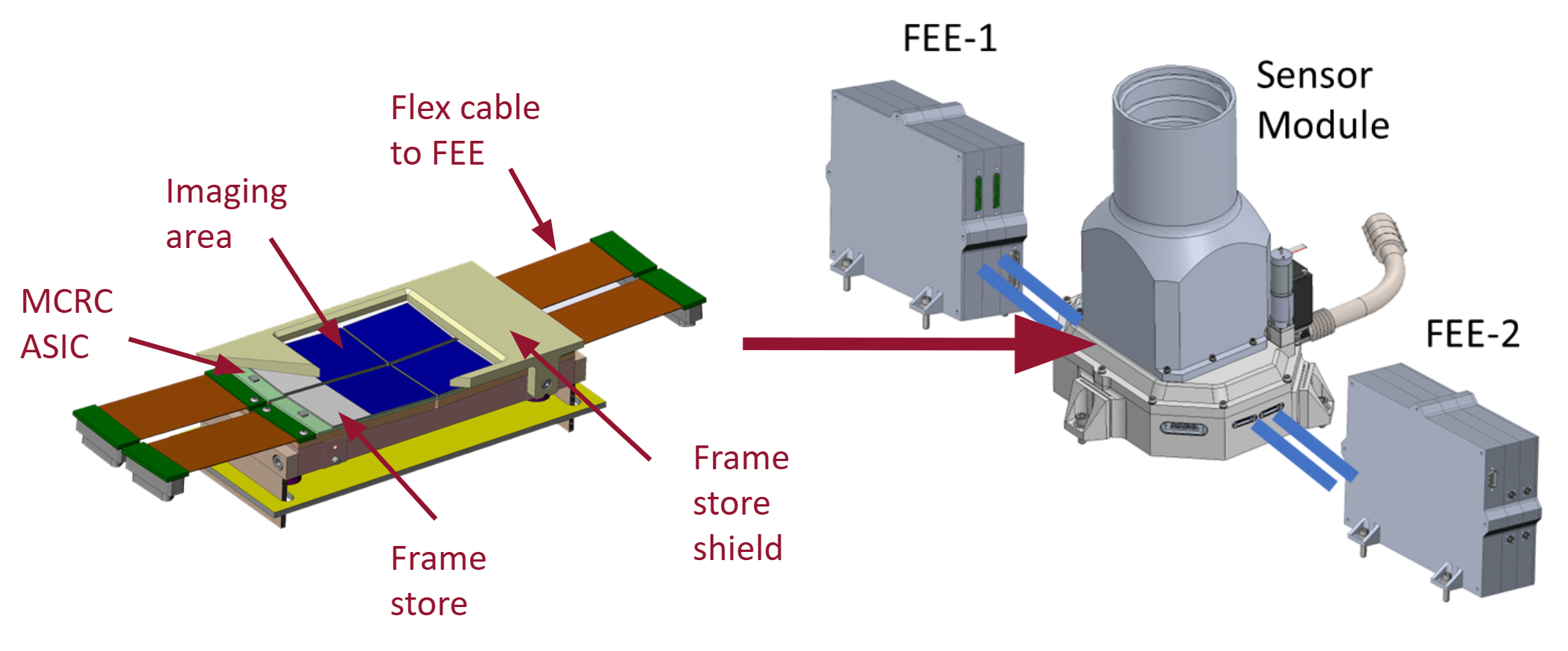}
    \caption{The AXIS focal plane assembly and sensor module.}
    \label{fig:axis-camera}
\end{figure}

The focal plane assembly of the AXIS camera, shown in Fig. \ref{fig:axis-camera}, contains an array of four 1440×1440 pixel CCDs, each with 16 parallel output channels.\cite{eric25-axiscamera} MIT-LL has fabricated prototype AXIS CCDs, designated CCID-100s, which MIT and Stanford have been commissioning\cite{stueber26-ccid100, lamarr26-ccid100}. Each CCD has its own independent readout chain consisting of an MCRC readout ASIC and an FEE unit.

The AXIS FEE captures the 16 differential channels from the MCRC, computes pixel values, and transmits those values to the BEE over a 1 Gbit Ethernet connection for event reconstruction. The selected ADCs (MCP37D31-200) can sample at 200 Msps shared across up to 8 channels; with 4 ADCs, the FEE can sample each channel at 50 Msps. The FPGA processing architecture provides flexibility in the CCD operating point: the pixel clock frequency and frame rates are flexible and can be chosen according to the desired tradeoff between noise performance and speed. On the high-speed end, for example, the CCD can operate with a 3.5 MHz serial clock at 20 frames per second. For this configuration, the FPGA ingests 14 samples per pixel, and on the output side, the 1 Gbit Ethernet link provides sufficient bandwidth: the payload data rate is only $$ 20\,\frac{\text{frames}}{\text{s}} \cdot (1440 \cdot 1440)\,\frac{\text{pixels}}{\text{frame}} \cdot 16\,\frac{\text{bits}}{\text{pixel}} \approx 664 \,\text{Mbps} $$ which is well within the 1 Gbit line rate. This leaves bandwidth for UDP/IP overhead and optional metadata such as sequence numbers. 

For improved noise performance, a slower pixel clock and frame rate can be chosen, which the FEE can support just as well. In this case, the FPGA will be able to take more samples per pixel, and less of the Ethernet bandwidth will be used. In the following section, a 2 MHz pixel clock is used as a baseline condition.


\section{FPGA DESIGN AND VALIDATION}

\begin{figure}[t!]
    \centering
    \includegraphics[width=.8\linewidth]{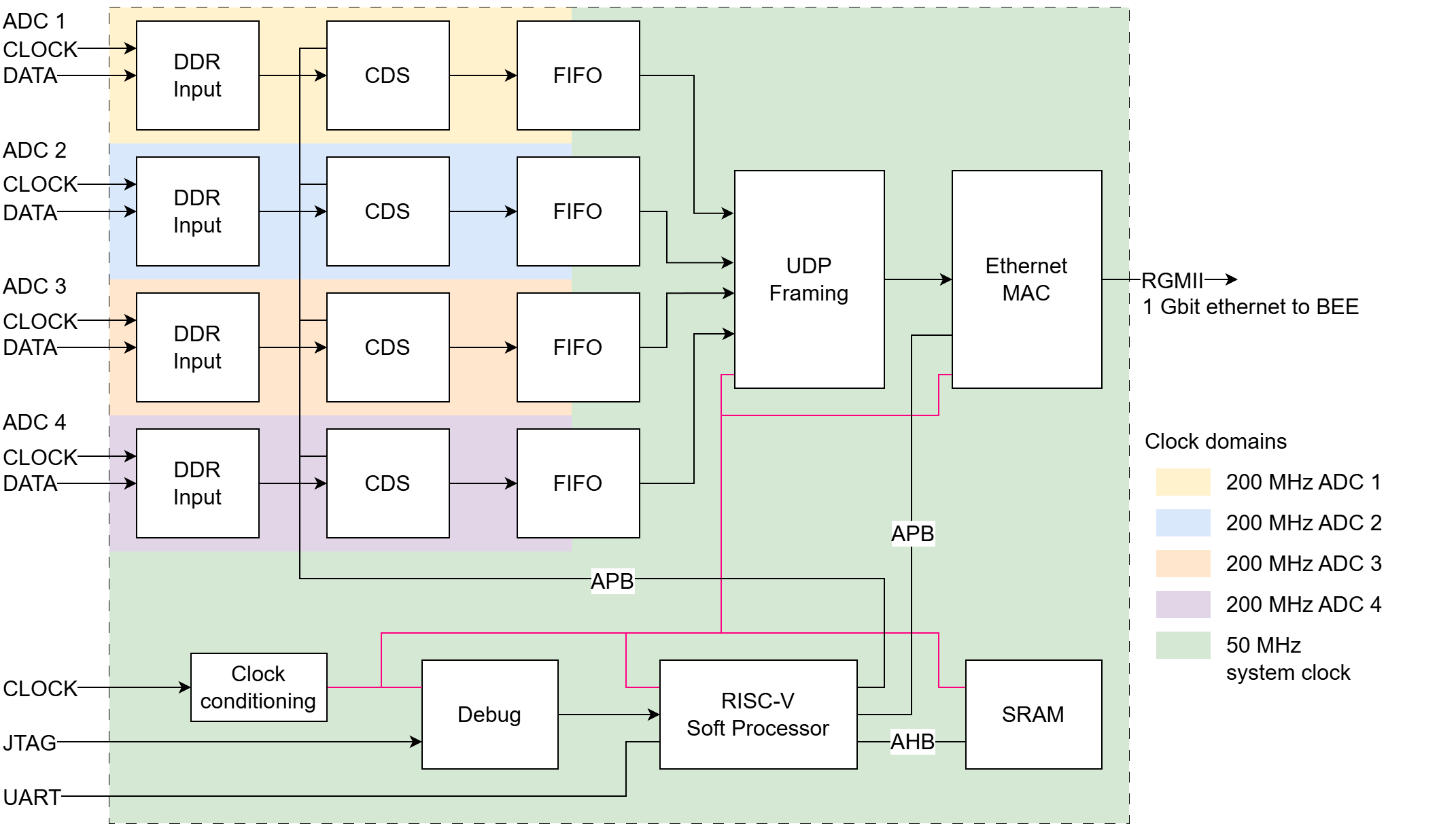}
    \caption{The FPGA fabric modules, clock domains, and the flow of data. The four sampling and CDS clock domains are clocked by each respective ADC sampling clock. Dual-clock FIFOs are employed to buffer pixel data and cross to the main system clock domain, which is clocked by an on-board oscillator. The UDP framer streams packets to the Ethernet MAC consisting of a UDP/IP header, pixel data from each FIFO, and an application-layer CRC32. The Mi-V soft processor executes instructions from a region of block SRAM.}
    \label{fig:fpga-fabric}
\end{figure}

The FPGA ingests digitized data from the four ADCs and performs CDS to recover pixel values. Pixel values are buffered in FIFOs\footnote{https://www.microchip.com/en-us/products/fpgas-and-plds/ip-core-tools/corefifo} (first-in, first-out) to cross the clock domain from each ADC's 200 MHz sampling clock to the 50 MHz system clock (Fig. \ref{fig:fpga-fabric}). Data are framed into UDP packets and forwarded to the Ethernet MAC\footnote{https://www.microchip.com/en-us/products/fpgas-and-plds/ip-core-tools/coretse} for transmission to the BEE. A Mi-V\footnote{https://www.microchip.com/en-us/products/fpgas-and-plds/system-on-chip-fpgas/mi-v/soft-cpus} soft processor exposes a serial command interface to the user for configuring components and controlling data flow. In particular, the processor configures the ADCs over their respective Serial Peripheral Interface (SPI) buses, configures the Ethernet PHY, and enables or disables the data stream. Through the command interface, a user can also select ADC test patterns and inspect a small shared memory region used by the CDS engine to verify data formatting and integrity. A debug module\footnote{https://www.microchip.com/en-us/products/fpgas-and-plds/ip-core-tools/corejtagdebug} enables line-by-line execution control for the processor.

\begin{figure}[t!]
    \centering
    \includegraphics[width=.8\linewidth]{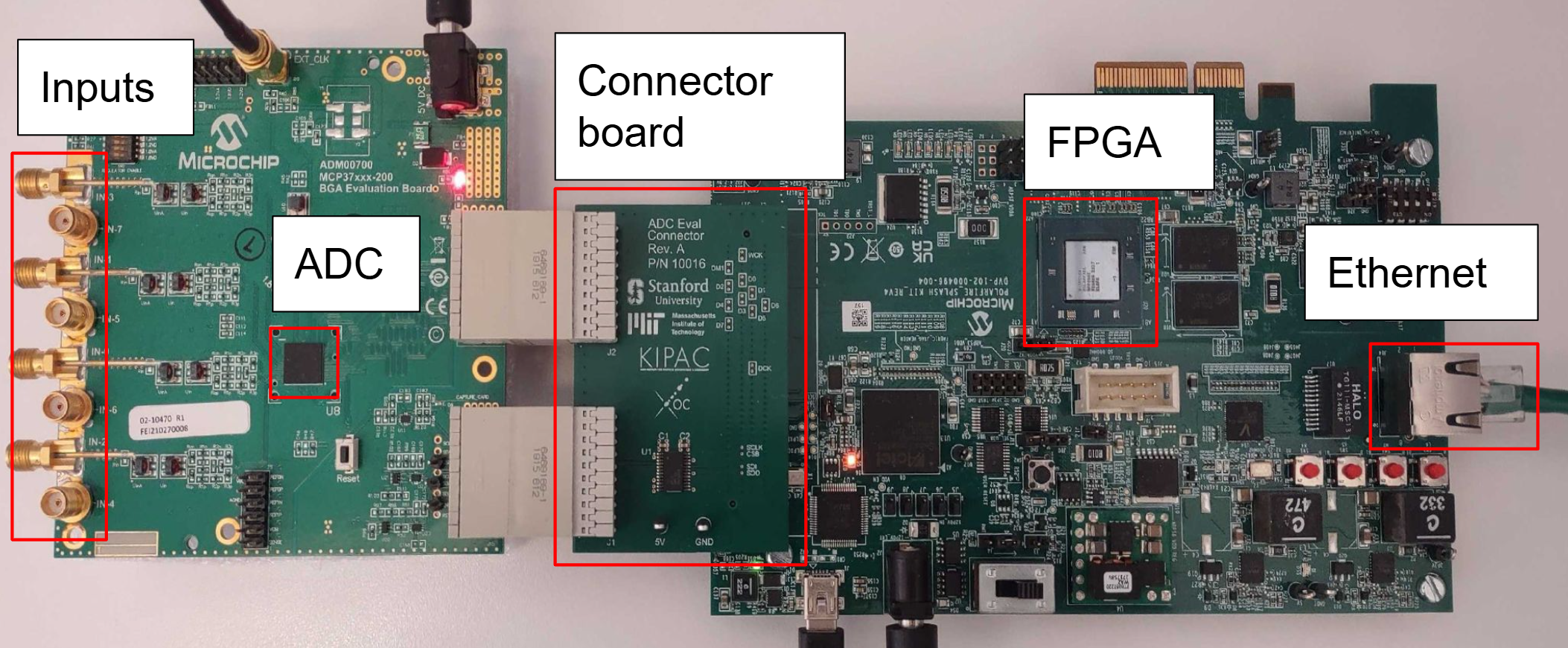}
    \caption{FPGA development hardware: on the left is an ADM00700 evaluation board for a MCP37D31-200 ADC and on the right is a PolarFire Splash Kit evaluation board for a MPF300T FPGA. They are linked by a custom connector board that routes the ADC SPI bus and low-voltage differential signaling (LVDS) data and clock signals to the FPGA.}
    \label{fig:test-hardware}
\end{figure}

As a proof of concept, we built the hardware platform shown in Fig. \ref{fig:test-hardware} to validate the FPGA design. The hardware platform is composed of evaluation boards\footnote{https://www.microchip.com/en-us/development-tool/adm00700}\footnote{https://www.microchip.com/en-us/development-tool/mpf300-splash-kit} from Microchip for the selected ADC and FPGA bridged by a custom connector board. Although this platform includes only a single ADC, the remaining channels are emulated with dummy-data generators so that processing, buffering, and throughput can be tested under representative conditions.

\begin{figure}[t!]
    \centering
    \includegraphics[width=.4\linewidth]{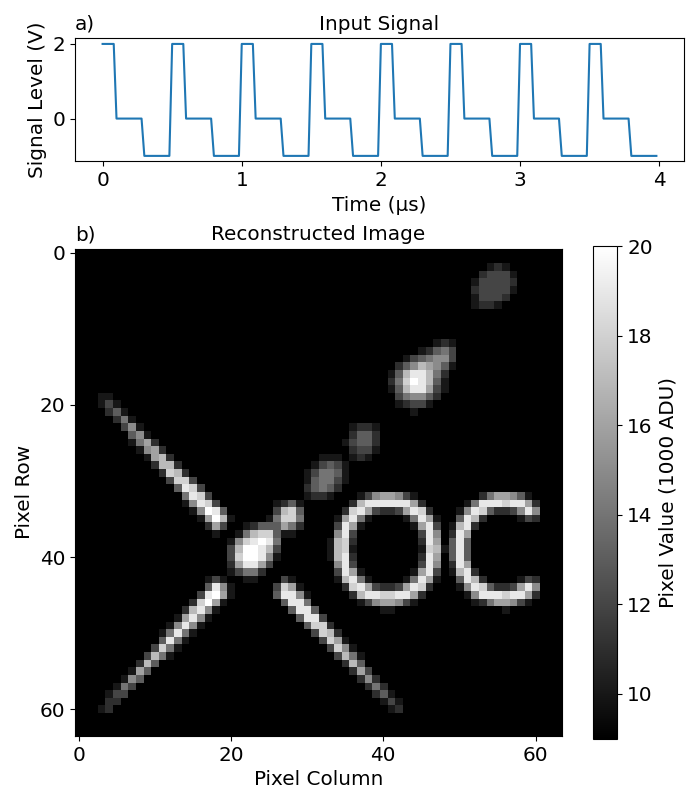}
    \caption{Sample readout for a single channel. The input signal shown in (a) is a 2 MHz CCD-like pulse with a reset spike, signal, and baseline. This signal is injected to the ADC input with a Keysight 33600A waveform generator. The FPGA samples the input at 50Msps and performs CDS, subtracting the average baseline level from the average signal level, to produce the pixel values shown in (b).}
    \label{fig:pixel-values}
\end{figure}

By driving a CCD-like waveform with known signal levels into one channel, we can verify that the CDS engine functions properly (Fig. \ref{fig:pixel-values}). In this test, an input signal is injected that mimics a CCD output channel, with signal levels for each pixel chosen according to a test image. The injected signal includes a frame sync pulse at the start of the image which the FPGA recognizes. After the frame sync, the timing for the rest of the image readout is deterministic (as it is with an operational CCD), so a simple finite state machine (FSM) tracks the image readout timing with no additional input. The CDS engine uses the timing information from the FSM to add and subtract the appropriate samples for each pixel and clocks out finished pixel values to a FIFO as shown in Fig. \ref{fig:fpga-fabric}. By capturing the UDP packets on the receiving end of the Ethernet connection and reconstructing the image (Fig. \ref{fig:pixel-values}), we verify that the FPGA correctly read out the desired test image.

This initial demonstration validates several high-risk elements of the architecture before committing to a full 16-channel design: high-speed LVDS capture, clock-domain crossings, CDS computation, and sustained packet streaming. By emulating the additional channels with dummy data generators, we test the internal bandwidth and buffering behavior under realistic aggregate rates while keeping the physical hardware simple.


\section{FUTURE WORK}

To fully validate the embedded readout system, we plan to operate it with our X-ray beamline in parallel with our existing readout system. The AXIS Testing and Acquisition Platform (AXIS-TAP) will enable these tests.\cite{juneau26-axis-tap} The AXIS-TAP board includes 4 ADCs and a daughter FPGA board from Brookhaven National Laboratory, enabling full 16-channel readout validation. The platform can be connected to our existing X-ray beamline,\cite{stueber2024,pan2025_spie_beamline} allowing parallel readout of live X-rays through this new embedded system and our commercial Archon CCD controller. This parallel readout will facilitate validation of the new system's accuracy and enable comparison of performance metrics in planned follow-up tests.

The next steps after achieving the desired readout behavior with an actual detector would be to generate the clocks and biases necessary to operate the CCD from the same platform. Developing a complete embedded CCD controller and readout system would represent an important step forward in developing the technologies necessary for future high-throughput X-ray observatories. 


\section{CONCLUSION}

The Stanford XOC team, in collaboration with MKI and MIT-LL, is developing high-speed readout electronics to enable the AXIS CCD to operate at the frame rates required to mitigate pile-up. We have implemented an FPGA-based FEE architecture that receives ADC samples, performs CDS, and streams pixel data over a 1 Gbit Ethernet connection to the backend computer. A proof-of-concept hardware demonstration validates the end-to-end data path and CDS functionality under representative throughput conditions. 

Near-term work will scale the design to full 16-channel operation on the AXIS-TAP board and will incorporate live CCD and X-ray testing to characterize noise and spectral performance relative to our established readout system. These developments provide a path towards an embedded CCD readout solution suitable for future high-throughput X-ray missions.

\acknowledgments 

This work has been supported by the NASA \textit{Astrophysics Probe Explorer} (APEX) program under contract number 80GSFC25CA019 and the NASA \textit{Strategic Astrophysics Technology} (SAT) program under grant number 80NSSC23K0211.


\bibliography{reference}   
\bibliographystyle{spiebib}   

\end{document}